# NUMERICAL STUDIES OF CO2 LEAKAGE REMEDIATION BY MICP-BASED PLUGGING TECHNOLOGY

**D. Landa-Marbán[1*], K. Kumar[2], S. Tveit[1], S.E. Gasda[1]**

[1] NORCE Norwegian Research Centre AS, Bergen, Norway

[2] University of Bergen, Bergen, Norway

* Corresponding author e-mail: dmar@norceresearch.no

## Abstract

Microbially induced calcite precipitation (MICP) is a technology for sealing leakage paths to ensure the safe storage of $CO_2$ in geological formations. In this work we introduce a numerical simulator of MICP for field-scale studies. This simulator is implemented in the open porous media (OPM) framework. We compare the numerical results to simulations using an upgraded implementation of the mathematical model in the MATLAB® reservoir simulation toolbox (MRST). Finally, we consider a 3D system consisting of two aquifers separated by caprock with a leakage path across the width of the reservoir. We study a strategy where microbial solution is injected only at the beginning of the treatment and subsequently either growth solution or cementation solution is injected for biofilm development or calcite precipitation. By applying this strategy, the numerical results show that the MICP technology could be used to seal these leakage paths.

## 1. Introduction

Carbon capture and storage (CCS) in the subsurface is one of the most promising scalable technologies for reduction of greenhouse gases through large scale carbon sequestration. For the successful implementation of this technology, the captured carbon must remain stored and any possibility of its migration back to the surface through the complex subsurface geometric structures such as faults, fractures, and abandoned wells must be reduced. Geological sequestration of $CO_2$ involves the injection of $CO_2$ into underground formations such as oil-bearing formations and saline aquifers allowing it to be trapped by the caprocks as has been practically achieved in the Norwegian continental shelf, e.g., in the Sleipner field. Many investigations have been conducted on assessment of $CO_2$ leakage (see e.g., [1-4]). The migration of $CO_2$ either to the freshwater aquifer or to the surface threatens not just the viability of the technology but also poses a risk to the precious ground water resource. This underlines the need for developing technologies that ensure closure of any potential leakage pathway for the trapped $CO_2$. Fig. 1 shows a schematic representation of injection of $CO_2$ for storage in deep subsurface, where leakage paths in caprocks lead to contamination of fresh water.

Microbial induced calcite precipitation (MICP) is an in-situ sealing technology that utilizes the biochemical processes to create barriers by calcium carbonate cementation. It involves injection of a mix of components including microbes, growth medium, and other chemical substances into a reservoir where the microbes produce calcite that reduces the in-situ permeability reducing the chances of $CO_2$ leakage. The field studies indicate that this strategy works in practice; [5] showed that MICP under field conditions improved the strengthening of the soil and concluded that MICP can be used for large-scale applications. One of the potential MICP applications is on reservoirs where leakage mitigation is relevant before $CO_2$ is injected. If the location of these leakage paths is known, then an injection strategy could be applied to create barriers by calcium carbonate precipitation in the leakage paths. Besides the sealing technology for prevention of $CO_2$ leakage, the MICP also has potential applications in biomineralized concrete [6], wastewater treatment, and erosion control.

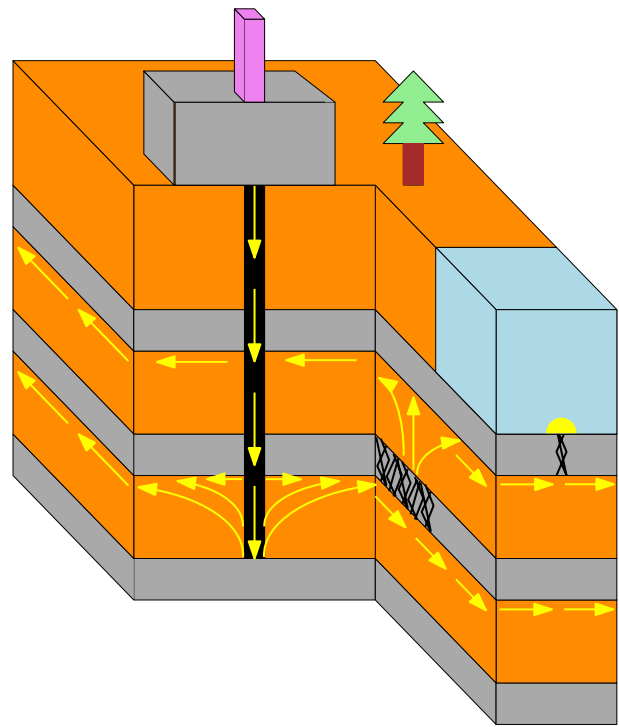

Figure 1: Contamination of water by $CO_2$ leakage.

The complexity and economics of the laboratory experiments at field scale necessitate the development of mathematical models that can enable simulation of the process. This development is built on the laboratory experiments conducted at the pore and core scales. Bai et al. [7] performed MICP experiments in microfluidic cells to study the calcite precipitation at the pore scale enabling the determination of certain coefficients in the mathematical model. The core-scale experiments by [8-10] help us identify the key processes involved. Based on these, detailed mathematical models have been proposed in literature. We refer to the model of [11] for a comprehensive pore-scale model for MICP and [12] for a comprehensive core-scale model. Based on the pore- and core-scale models, [13-14] developed simplified mathematical models for MICP at the field scale. An optimization study under parameter uncertainty using a simplified MICP model is presented in [15]. In this work, we will be relying on the model developed in [14].

Our motivation is to develop mathematical models and numerical tools that simulates the MICP at the field scale. The process is quite complex taking place at several scales, with the flow and transport in complex geometry, and involving a set of reactions at multiple temporal and spatial scales. This yields a system of coupled partial differential equations and a set of ordinary differential equations with heterogeneous coefficients, nonlinearities, and degeneracies, e.g., multiphase flows coupled to reactive transport. Performing numerical simulations of such a complex set of mathematical equations is quite difficult and often intractable even with the usual techniques of local grid refinement, multiscale approaches, or improving the time stepping. Another challenge in considering highly detailed models is the unavailability of the laboratory data at field scale to estimate the increasing number of parameters involved as the complexity of models is increased. We are thus motivated by the following consideration: we would like to consider models that are simple enough to let us perform simulations at the field scale, yet capture the essence of the physical and chemical processes keeping the number of parameters as small as we can. The simplicity of the model considered here will also allow us to perform optimization of the injection strategy.

In this proceeding, we consider a field-scale model for MICP and use this to study the remediation of a fractured zone in a caprock to prevent the potential leakage of $CO_2$. More specifically, we consider both 2D and 3D geometries with potential leakage pathways and inject microbes mixed in water, a growth medium, and cementation solution to allow the calcite precipitation. We then study the evolution of the in-situ porosity and the flux of the $CO_2$. This allows us to simulate the process and study its effectiveness as a sealing technology. In this work, we implement the mathematical model in the open porous media (OPM) framework which is an open-source tool for simulating multiphase flow and transport in subsurface porous media [16]. We compare the results with those from the proto-type model developed in MRST, a MATLAB® based reservoir simulation toolbox [17]. The use of open-source toolbox allows us also to benefit from the existing implementation of other modules of a reservoir simulator such as multiphase modules, and handling of complex geometry. Finally, we study a 3D system consisting of two aquifers separated by caprock with a leakage path across the width of the reservoir. We conclude with the discussions on the numerical findings and the further work that we plan to undertake.

## 2. Methodology

MICP is a complex bio-geochemical process resulting in mineralization of $CO_2$ in the form of calcite as a result of microbial metabolic activities. A rough description of the process consists of an impermeable biofilm formation where microorganisms produce an adhesive matrix of extracellular polymeric substance (EPS), microbes catalyze urea to produce ammonium ions and carbonate, the calcium-rich environment then reacts with the carbonate ions to produce calcite. We consider accordingly a set of unknown variables consisting of biofilm, water, calcite, microbes suspended in water, oxygen (acting as the electron acceptor), and urea. A simplified representation of the MICP reactions is given by

$$CO(NH_2)_2 + 2H_2O + Ca^{2+} \xrightarrow{urease} 2NH_4^+ + CaCO_3 \downarrow.$$

We consider the rate-limiting components in the injected solutions to be suspended microbes, oxygen, and urea respectively. We assume that the MICP treatment is applied before $CO_2$ injection. Thus, only water is presented in the aquifer. After MICP treatment, $CO_2$ is injected to the reservoir. Given these assumptions, we describe simplified models for MICP and $CO_2$ sequestration in the following subsections.

### 2.1 Mathematical model of MICP technology

A comprehensive presentation of the MICP model can be found in [14]. Here we shortly describe this mathematical model.

We consider a reservoir characterized by its initial porosity $\phi_0$ and permeability $\mathbb{K}_0$. We denote $\phi_b$ and $\phi_c$ as the volume fractions of biofilm and calcite respectively. Then the rock porosity reduction is

$$\phi = \phi_0 - \phi_b - \phi_c.$$

Since the biofilm and calcite are assumed immobile, the mass balance equations only involve the storage term,

$$\frac{\partial(\rho_\chi \phi_\chi)}{\partial t} = R_\chi, \quad \chi \in \{b, c\},$$

where $\rho_\chi$ are densities and $R_\chi$ reaction terms described later in this section.

We denote $\phi_{crit}$ as the value of porosity when the permeability reaches a minimum value $K_{min}$ ($\phi \leq \phi_{crit}$). The following relationship is used to model the permeability reduction as a consequence of the porosity reduction ($\phi_{crit} < \phi$)

$$\mathbb{K} = \left[\mathbb{K}_0 \left(\frac{\phi - \phi_{crit}}{\phi_0 - \phi_{crit}}\right)^\eta + K_{min}\right] \frac{\mathbb{K}_0}{\mathbb{K}_0 + K_{min}},$$

where $\eta$ is a fitting factor.

The mass conservation equation and Darcy's law for the water are

$$\frac{\partial \phi}{\partial t} + \nabla \cdot \boldsymbol{v_w} = q_w, \quad \boldsymbol{v_w} = -\frac{\mathbb{K}}{\mu_w}(\nabla p_w - \rho_w \boldsymbol{g}),$$

where $\boldsymbol{v_w}$ is the discharge per unit area, $\rho_w$ the fluid density, $p_w$ the reservoir pressure, $\boldsymbol{g}$ the gravity, $\mu_w$ the water viscosity, and $q_w$ the source/sink term.

The mass balances for the suspended microbes (m), oxygen (o), and urea (u) are

$$\frac{\partial(c_\xi \phi)}{\partial t} + \nabla \cdot (c_\xi \boldsymbol{v_w}) = c_\xi q_w + R_\xi, \quad \xi \in \{m, o, u\},$$

where $c_\xi$ are mass concentrations and $R_\xi$ reaction terms described next.

The reaction term for the suspended microbes is

$$R_m = c_m \phi \left( Y\mu \frac{c_o}{k_o + c_o} - k_d - k_a \right) + \phi_b \rho_b k_{str} \phi \|\nabla p_w - \rho_w \boldsymbol{g}\|^{0.58},$$

where $Y$ is the growth yield coefficient, $\mu$ is the maximum specific growth rate, $k_o$ is the half-velocity coefficient of oxygen, $k_d$ is the microbial death coefficient, $k_a$ is the microbial attachment coefficient, and $k_{str}$ is the detachment rate.

The reaction term for the oxygen is

$$R_o = -(c_m \phi + \rho_b \phi_b) F\mu \frac{c_o}{k_o + c_o},$$

where $F$ is the mass ratio of oxygen consumed to substrate used for growth.

The reaction term for the urea is

$$R_u = -\rho_b \phi_b \mu_u \frac{c_u}{k_u + c_u},$$

where $\mu_u$ is the maximum rate of urea utilization and $k_u$ is the half-velocity coefficient for urea.

The reaction term for the biofilm is

$$R_b = \rho_b \phi_b \left[ Y\mu \frac{c_o}{k_o + c_o} - k_d - \frac{R_c}{\rho_c(\phi_0 - \phi_c)} - k_{str} \phi \|\nabla p_w - \rho_w \boldsymbol{g}\|^{0.58} \right] + c_m \phi k_a.$$

The reaction term for the calcite is

$$R_c = -\rho_b \phi_b Y_{uc} \mu_u \frac{c_u}{k_u + c_u},$$

where $Y_{uc}$ is a yield coefficient (units of produced calcite over units of urea utilization).

## 2.2 Mathematical model of $CO_2$ storage

For simulation of $CO_2$ sequestration, we consider a simple immiscible two-phase flow model. We denote water saturation as $s_w$ and $CO_2$ saturation as $s_{CO_2}$, where $s_w + s_{CO_2} = 1$. The mass conservation and extended Darcy's law for each $\alpha$ phase ($\alpha = w, CO_2$) are

$$\phi \frac{\partial s_\alpha}{\partial t} + \nabla \cdot \boldsymbol{v_\alpha} = q_\alpha, \quad \boldsymbol{v_\alpha} = -\frac{k_{r,\alpha}}{\mu_\alpha} \mathbb{K}(\nabla p_\alpha - \rho_\alpha \boldsymbol{g}),$$

where $k_{r,\alpha}$ are relative permeabilities. The relative permeabilities are set as a linear function of the saturations ($k_{r,\alpha} = s_\alpha$) and the capillary pressure is neglected ($p_{CO_2} = p_w$).

## 2.3 Implementation

The open porous media (OPM) initiative, a free open-source software for reservoir modeling and simulation, is used to implement the MICP mathematical model [16]. The source code for OPM and its related modules can be obtained at http://github.com/OPM and a description of the simulator can be found in the OPM Flow manual [18]. The implementation of this MICP model is made in the 2020.10 release of OPM. The source code for the implementation of this MICP model in MRST can be obtained at https://www.sintef.no/projectweb/mrst/ and https://github.com/daavid00/ad-micp.git. We have upgraded the first version of the ad-micp module to make it compatible with GNU Octave.

The MICP model is solved on domains with vertex-centered grids. The leakage paths are discretized with three-dimensional elements. This approach has shown to be in good agreement with comparable simulations using different approaches such as discrete fracture networks (see e.g., [19]). Two-point flux approximation (TPFA) and backward Euler (BE) are used for the space and time discretization respectively. The resulting system of equations is linearized using the Newton-Raphson method. We consider constant-pressure production wells on the domain boundaries to model an infinite acting aquifer. We use the MRST functionalities to produce the grids and write the corresponding grid file in GRDECL format which is read by the OPM simulator. We use OPM Flow to assess the $CO_2$ leakage.

## 2.4 Model parameters and reservoir properties

Here we have separated model inputs in two parts. Table 1 shows the model parameters regarding the fluid properties and MICP processes. These values are selected from previous studies and full references to the sources can be found in [14].

Table 1: Table of model parameters for the numerical studies.

| Parameter | Sym | Value | Unit |
|---|---|---|---|
| Density (biofilm) | $\rho_b$ | 35 | $kg/m^3$ |
| Density (calcite) | $\rho_c$ | 2710 | $kg/m^3$ |
| Density ($CO_2$) | $\rho_{CO_2}$ | 479 | $kg/m^3$ |
| Density (water) | $\rho_w$ | 1045 | $kg/m^3$ |
| Detachment rate | $k_{str}$ | $2.6 \times 10^{-10}$ | $m/(Pa\ s)$ |
| Half-velocity coefficient (oxygen) | $k_o$ | $2 \times 10^{-5}$ | $kg/m^3$ |
| Half-velocity coefficient (urea) | $k_u$ | 21.3 | $kg/m^3$ |
| Maximum specific growth rate | $\mu$ | $4.17 \times 10^{-5}$ | $1/s$ |
| Maximum rate of urea utilization | $\mu_u$ | $1.61 \times 10^{-2}$ | $1/s$ |
| Microbial attachment rate | $k_a$ | $8.51 \times 10^{-7}$ | $1/s$ |
| Microbial death rate | $k_d$ | $3.18 \times 10^{-7}$ | $1/s$ |
| Oxygen consumption factor | $F$ | 0.5 | [−] |
| Viscosity ($CO_2$) | $\mu_{CO_2}$ | $3.95 \times 10^{-5}$ | $Pa\ s$ |
| Viscosity (water) | $\mu_w$ | $2.54 \times 10^{-4}$ | $Pa\ s$ |

| Parameter | Sym | Value | Unit |
| --- | --- | --- | --- |
| Yield coefficient (growth) | $Y$ | 0.5 | [−] |
| Yield coefficient (calcite/urea) | $Y_{uc}$ | 1.67 | [−] |

Table 2 shows the reservoir properties we consider for the simulations. These properties are set to common values for numerical studies.

Table 2: Table of reservoir properties for the simulations.

| Parameter | Sym | Value | Unit |
| --- | --- | --- | --- |
| Aperture (leak) | $a$ | 1 | $m$ |
| Fitting factor | $\eta$ | 3 | [−] |
| Gap (lower aquifer) | $g_l$ | 15 | $m$ |
| Gap (upper aquifer) | $g_u$ | 5 | $m$ |
| Height (aquifer) | $H$ | 5 | $m$ |
| Height (caprock) | $h$ | 20 | $m$ |
| Length (aquifer) | $L$ | 100 | $m$ |
| Gap (potential leakage) | $l$ | 15 | $m$ |
| Permeability (aquifer) | $\mathbb{K}_A$ | $10^{-14}$ | $m^2$ |
| Permeability (leakage) | $\mathbb{K}_L$ | $2\mathbb{K}_A$ | $m^2$ |
| Permeability (minimum) | $K_{min}$ | $10^{-20}$ | $m^2$ |
| Porosity (aquifer/leakage) | $\phi_0$ | 0.15 | [−] |
| Porosity (critical) | $\phi_{crit}$ | 0.1 | [−] |
| Tilt angle (leak) | $\theta$ | 135 | ° |
| Width (aquifer) | $W$ | 20 | $m$ |
| Width (leak) | $w$ | 6 | $m$ |

*2.5 Injection strategy*

Both laboratory experiments and numerical studies have shown that by separating the injection of solutions (microbial, growing, and cementation solutions) with no-flow periods then limited clogging is expected to occur near the injection site (see e.g., [8,12]). We adopt this strategy, where first the microbial solution is injected ($t_1^I$) and after only water is injected to displace the microbes deeper in the aquifer ($t_2^I$). After closing the system to allow the suspended microbes to attach themselves to the rock ($t_3^I$), growth solution is injected ($t_4^I$) to stimulate the biofilm formation, followed by only water injection ($t_5^I$) and a non-flow period ($t_6^I$). Next, the cementation solution is injected ($t_7^I$) followed by only water ($t_8^I$) and a non-flow period ($t_9^I$) to allow the calcite precipitation to occur. All these nine injection times are denoted as phase I. Several phases can be applied to seal the leak.

For the numerical studies we fix the injected concentrations, and we keep the injection rate constant in each of the phases. The injected concentrations are set to $c_m = 0.01\ kg/m^3$, $c_o = 0.04\ kg/m^3$, and $c_u = 300\ kg/m^3$. The value of injection rate $Q_w$ is different for the different flow systems and its value is given in each example.

## 3. Results and Discussion

In this section, three examples are presented to demonstrate the validity and application of the implementation in OPM. The first example verifies the simulator against a previous implementation in MRST. In the second example, we show the evolution of the MICP processes during injection of one MICP treatment on a 2D flow domain with a diagonal leakage path. In the last example, we describe a successful injection strategy to mitigate $CO_2$ leakage on a reservoir with a leakage path along the width of the caprock.

*3.1 Example 1: comparison of the model implementation in OPM to the one in MRST*

The comparison between implementations is performed on a 1D flow horizontal system as shown in Fig. 2.

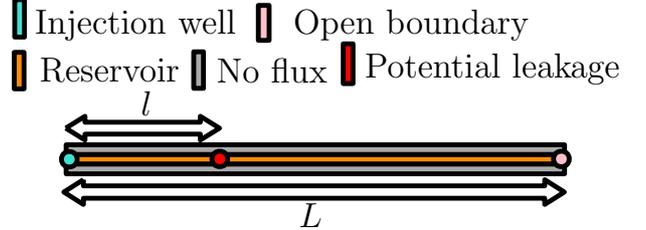

Figure 2: 1D domain with a potential leakage zone.

The injection well is located on the left and the production well on the right side. The potential leakage zone is located at $[12.5\ m, 17.5\ m]$ from the injection well. The size of the domain is $100 \times 1 \times 1\ m$ and the dimension of the grid is $100 \times 1 \times 1$. The injection rate is set to $2.31 \times 10^{-5}\ m^3/s$. Fig. 3 shows the simulation results obtained from both implementations.

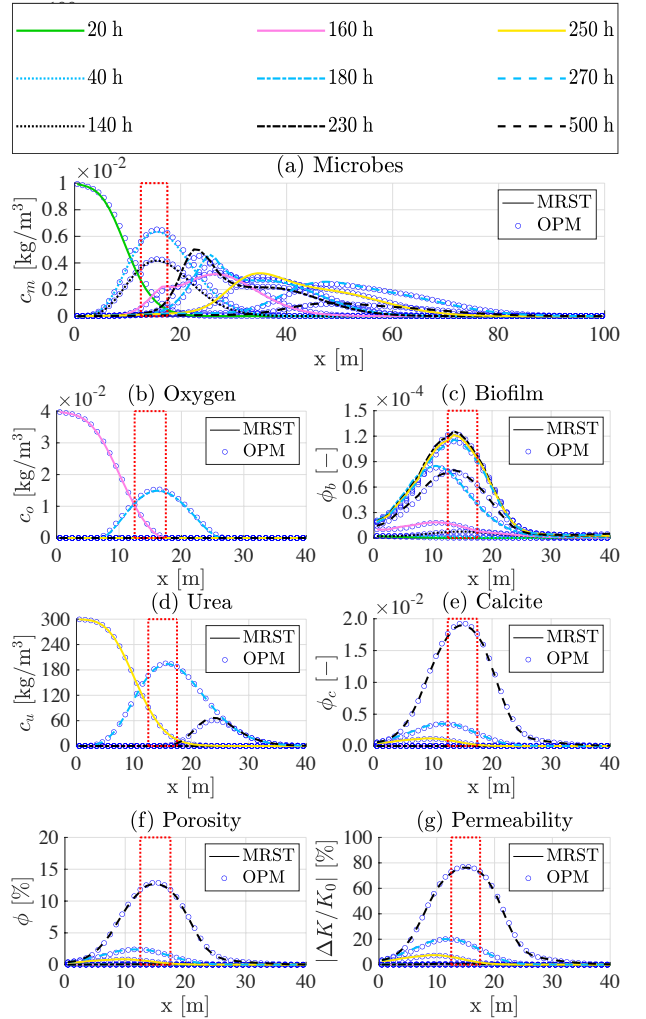

Figure 3: Comparison between implementations in OPM and MRST.

As shown in Fig. 3, the results verify that there is a good agreement in the spatial distribution of the model variables over time computed by OPM and MRST. For this example, the simulation time using OPM (2.82 $s$) is ca. 20 times faster than using MRST (55.81 $s$).

*3.2 Example 2: MICP on a 2D flow domain with a diagonal leak*

To demonstrate the use of the MICP simulator in the presence of a leakage path, we consider the 2D flow system shown in Fig. 4.

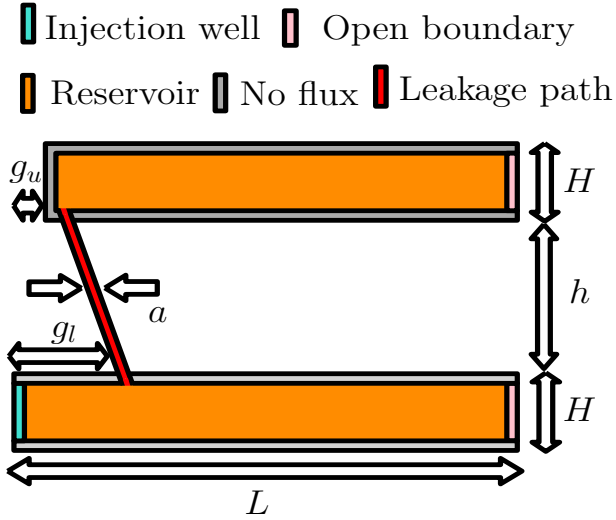

Figure 4: 2D domain with a leakage path.

The times for the injection strategy are $t_1^I = 15\,h$, $t_2^I = 22\,h$, $t_3^I = 100\,h$, $t_4^I = 130\,h$, $t_5^I = 135\,h$, $t_6^I = 160\,h$, $t_7^I = 200\,h$, $t_8^I = 210\,h$, and $t_9^I = 300\,h$ and the injection rate is set to $2.31 \times 10^{-4}\,m^3/s$. Fig. 5 shows the spatial discretization and simulation results after application of one phase of MICP treatment.

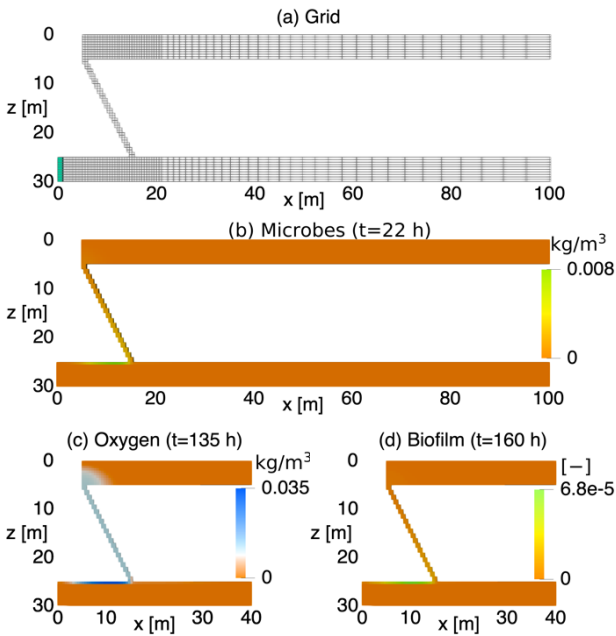

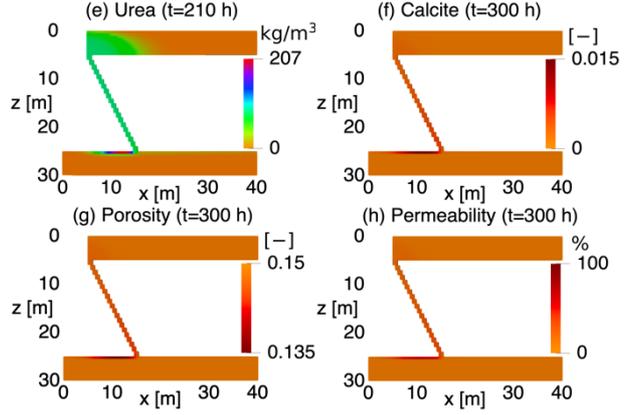

Figure 5: (a) Grid with well and spatial distribution of (b) suspended microbes, (c) oxygen, (d) biofilm, (e) urea, (f) calcite, (g) porosity, and (f) permeability reduction at different times of the MICP treatment.

In Fig. 5h we observe a significant permeability reduction on the leakage path (max ca. 30%) after only one phase of MICP treatment.

*3.3 Example 3: leakage mitigation on a 3D flow domain with a diagonal leak*

The implementation of the MICP mathematical in OPM allows for computationally challenging simulations. We consider the 3D system shown in Fig. 6.

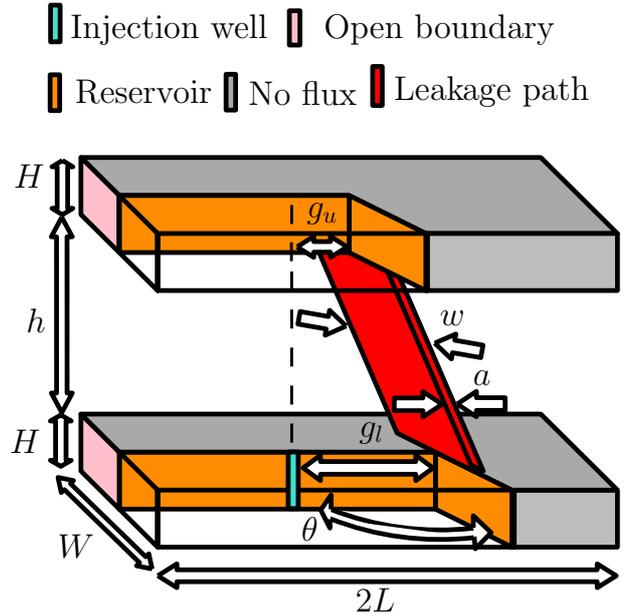

Figure 6: 3D domain with a leakage path.

We proceed to design an injection strategy for the sealing of the leakage path. Here we use an ad-hoc approach where we run simulations and change manually the injection times and rates. The following times and rates result in the successful sealing of the leakage path after five phases of MICP treatment: $Q_w^I = Q_w^{II} = Q_w^{III} = 8.70 \times 10^{-3}\,m^3/s$, $t_1^I = 15\,h$, $t_2^I = 22\,h$, $t_3^I = 100\,h$, $t_4^I = 130\,h$, $t_5^I = 135\,h$, $t_6^I = 160\,h$, $t_7^I = 200\,h$, $t_8^I = 210\,h$, $t_9^I = 300\,h$, $t_4^{II} = 330\,h$, $t_5^{II} = 340\,h$, $t_6^{II} = 341\,h$, $t_7^{II} = 371\,h$, $t_8^{II} = 381\,h$, $t_9^{II} = 431\,h$, $t_7^{III} = 461\,h$, $t_8^{III} = 471\,h$, $t_9^{III} = 571\,h$, $t_4^{IV} = 601\,h$, $t_5^{IV} = $

611 $h$, $t_6^{IV} = 612\ h$, $t_7^{IV} = 642\ h$, $t_8^{IV} = 652\ h$, $t_9^{IV} = 702\ h$, $t_7^V = 732\ h$, $t_8^V = 742\ h$, and $t_9^V = 800\ h$. In this injection strategy microbial solution is injected only at the beginning of the treatment and subsequently either growth solution or cementation solution is injected for biofilm development or calcite precipitation. The grid and simulation results are shown in Fig. 7.

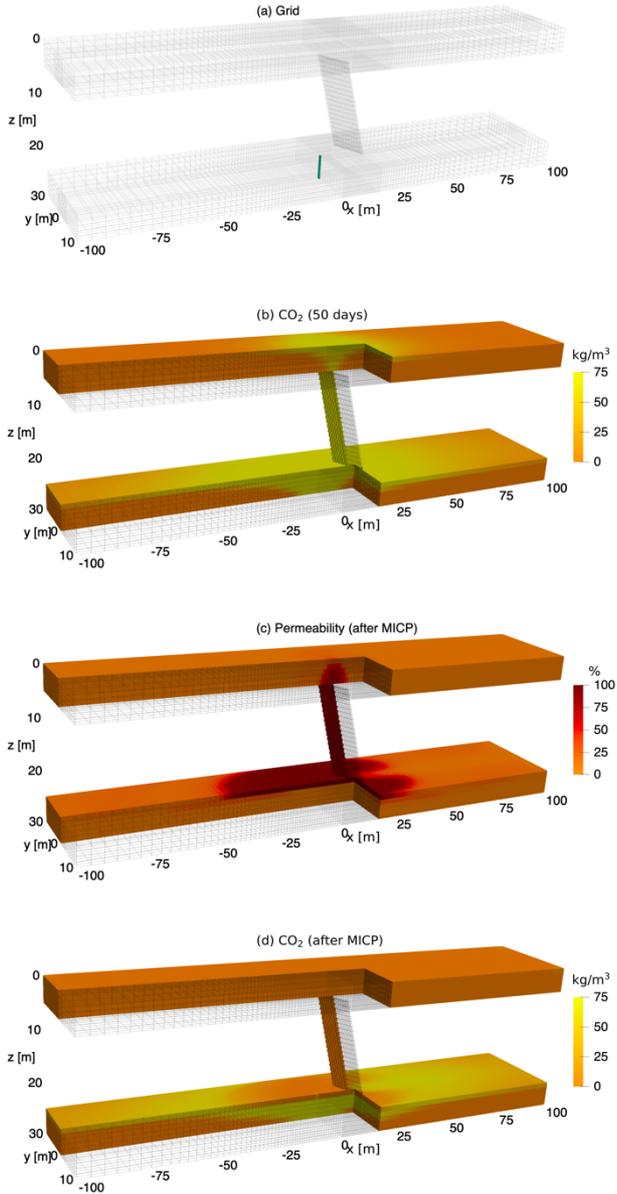

Figure 7: (a) Grid with well and spatial distribution of (b) $CO_2$ without MICP treatment, (c) permeability reduction, and (d) $CO_2$ after MICP treatment.

For a better visualization of the $CO_2$ rate through the leakage path, we plot the average normalized value of $CO_2$ flux along the width of the leakage path over time with and without MICP treatment in Fig. 8. The $CO_2$ was injected at a rate of $2.31 \times 10^{-4}\ m^3/s$.

We observe from the plot without MICP treatment that after few days of injection the $CO_2$ reaches the leakage path, and a significant amount of $CO_2$ leak to the upper aquifer. This leak is mitigated after application of the MICP technology.

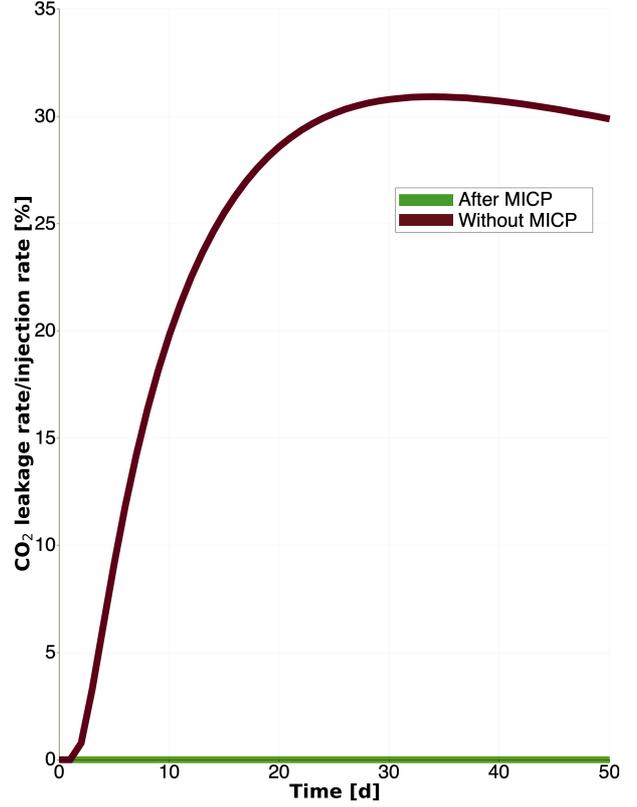

Figure 8: Leakage rate of $CO_2$ through the leakage path (at z=5 m) in the caprock.

## 4. Conclusions

This paper presents the application of the MICP technology for $CO_2$ leakage remediation using the OPM simulator. The implementation of the MICP mathematical model in OPM is compared against an implementation in MRST resulting in a good agreement between numerical results. Subsequently the application of the MICP simulator in OPM is demonstrated in a 2D flow domain with a leakage path. Finally, we design an injection strategy to seal a leakage path in a complex 3D system with a diagonal leakage path along the width of the caprock. This study demonstrates that it is possible to use MICP technology to plug a leakage pathway across the width of the reservoir.

Currently we are focusing on adding dispersion effects and making the implementation of the MICP model a part of OPM Flow, which in turn will make the model available as open-source code. Further work is to use this implementation to perform optimization and sensitivity analysis studies.

## Acknowledgements

The authors are grateful for the financial support from Research Council of Norway through the project "Efficient models for microbially induced calcite precipitation as a seal for $CO_2$ storage (MICAP)" (grant 268390).